# Using Conversational Agents To Support Learning By Teaching

**Nalin Chhibber**
David R Cheriton School of Computer Science, University of Waterloo
Waterloo, Canada
nalin.chhibber@uwaterloo.ca

**Edith Law**
David R Cheriton School of Computer Science, University of Waterloo
Waterloo, Canada
edith.law@uwaterloo.ca

## ABSTRACT

Conversational agents are becoming increasingly popular for supporting and facilitating learning. Conventional pedagogical agents are designed to play the role of human teachers by giving instructions to the students. In this paper, we investigate the use of conversational agents to support the 'learning-by-teaching' paradigm where the agent receives instructions from students. In particular, we introduce Curiosity Notebook: an educational application that harvests conversational interventions to facilitate students' learning. Recognizing such interventions can not only help in engaging students within learning interactions, but also provide a deeper insight into the intricacies involved in designing conversational agents for educational purposes.

**ACM Reference Format:**
Nalin Chhibber and Edith Law. 2019. Using Conversational Agents To Support Learning By Teaching. In . ACM, Glasgow, Scotland, UK, 7 pages. https://doi.org/10.1145/1122445.1122456

## INTRODUCTION

Learning is an important aspect of knowledge acquisition, but conventional pedagogical practices provide little motivation to develop "usable knowledge", and tend to promote memorization tactics over understanding. In recent years, many pedagogical tools have been introduced to deliver important lessons and provide deeper understanding of a topic. However, most of them focus on imitating prevalent instructional teaching techniques and do not place much emphasis on learning from social interaction, which is common in real life. Humans learn new things outside their regular classroom environment while engaging in conversation and discussion with friends, family, or peers. Thus, using conversational agents as intelligent tutors opens the door to myriad opportunities that enable learning through social interactions. An appropriately structured conversational agent can overcome the limitations of graphical interfaces and utilize a natural communication channel to support learning, by holding an on-topic dialogue with humans. Such agents can ask directed questions and engage users by using verbal and non-verbal interventions. In some aspects, they can also go beyond natural human limits such as limited patience or attention span, or fatigue from prolonged interaction. In addition to their unlimited capacity for interaction, conversational agents can also provide graphical traces of their thinking.

The idea of using virtual agent technologies for education has been explored in prior work, but the focus has either been on the agent's knowledge or its delivery mechanism, and not on the nature of its interactions. In this work, we introduce Curiosity Notebook, an educational environment that allows users to communicate with a conversational agent in natural language while performing a learning task.

## RELATED WORK

### Technology enabled learning

Increasing demand for diverse educational needs and greater personalization of lessons is fueling the research behind technology-based support in education. One of the first attempts in this direction was the use of Intelligent Tutoring System (ITS), that was based on on social models of learning in one-on-one human tutoring [16]. ITS enabled individual interactions between tutors and tutees, providing great potential for personalized education. However, it did not bring the notion of agents within the learning context and considered learners as passive knowledge recipients. To overcome this, developers and educational researchers have made many further attempts to build systems that can interact with human users in a more natural way, by intelligently mimicking human behaviors within the context of a specific application or task domain.

**Human-Agent Interaction in learning systems**

Identifying the appropriate role for conversational agents in a learning context is an important factor to consider when defining their use in the domain of education. Agents have been used as peers, tutors (teaching agents), and learners (teachable agents) in computer-aided learning applications.

*Teaching Agents:* Teaching agents have the longest history of research and development. These agents act as tutors and play the role of human teachers to provide direct curriculum support through hints, tutorials, and supervision.

*Peer Agents:* Peer agents are meant to serve as learning companions for a student in order to promote peer-to-peer interactions. Although this approach is different than tutor-to-student interaction, the agents are often presented as more knowledgeable peers who guide the students along a learning trajectory.

*Teachable Agents:* Teachable Agents allow students to teach a less-knowledgeable AI system to help facilitate learning on the part of the student. In this approach, the agent takes the role of a novice and asks student-teachers to guide them along a learning path.

While most of the previous research in computer-mediated learning applications is focused on using agents as peers [9, 14], or tutors that play the role of a human-teacher [7, 10], our research is catered towards the scenario where these agents take the role of a less intelligent entity, allowing the students to teach [3, 4]. This approach is inspired from the Protégé effect, which demonstrates that learning for the sake of teaching others is more beneficial than learning for one's own self. Previous work in cognitive science and education research supports the presence of the Protégé effect in reciprocal teaching [13], peer-assisted tutoring [6], small-group interaction [15] and self-explanation [5]. Studies focused on the cognitive benefits of teaching suggest that preparing to teach may produce more organized cognitive structures than learning the material for oneself [2]. Biswas et al. has shown that expecting to teach others helps in self-reflection, builds a sense of responsibility, and is useful for meaningful structuring of information [3]. This has been confirmed in later studies that demonstrate the effectiveness of the Protégé effect for cognitive [12], meta-cognitive [11] and motor learning skills [8]. Despite this, use of conversational agents, and their individual personality traits that can enhance learning outcomes through natural language, have not been fully explored across all dimensions.

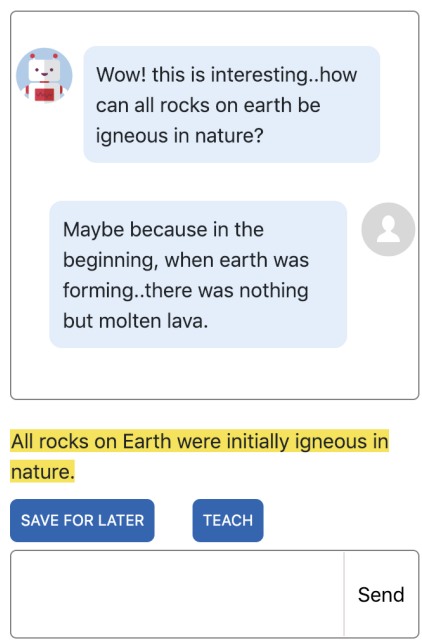

Figure 1: Flow of interaction between student and conversational agent. (a) student highlights a sentence from text, (b) agent asks the query, (c) student responds with the answer.

# CURIOSITY NOTEBOOK

Considering the problem space mentioned above, we propose a new learning environment in the form of *Curiosity Notebook* that supports the interaction between student-teachers and agent-learners. The purpose of this system is to focus on the nature of the interaction between tutor and tutee. Specifically, we aim to promote the scenario that enables students to take the role of an instructor who teaches an AI agent. Moreover, we are interested in finding the cognitive, affective, and meta-cognitive interventions that can help students learn new reasoning strategies while teaching. Through these conversational interventions, we intend to find useful characteristics of an agent's personality that can maximize student retention and overall engagement during a learning task. To support this, Curiosity Notebook allows students to select a topic and read related articles. While reading, students can highlight sentences and use them to direct conversations in natural language dialogues through a textual interface (Figure 1a). The agent then asks questions and prompts the student-teacher to elicit their queries and reveals what they do not understand about the topic (Figure 1b). These queries are answered by the student-teacher interacting with the agent (Figure 1c), allowing them to reflect upon their own knowledge and ultimately gain a better understanding about the topic. The system is built as an Angular web application with Python Flask in the back-end.

## Socio-Technical System

The socio-technical system associated with the Curiosity Notebook consists of the following three components: the learning environment, the student, and the conversational agent.

The **learning environment** consists of articles from different topics, and provides a mechanism for the student to communicate with the agent. It serves three primary purposes during the actual exchange of dialogue. First, it displays the learning progress of the agent as the student is teaching, providing an indication of when the discussion has deviated too much from the topic (e.g., when the student is teaching irrelevant information). Second, the environment applies sanity checks on the student's textual input by correcting spelling mistakes and converting complex sentence structures into simpler forms that are recognizable by the agent. This ensures that the information taught is consumable by the agent, and fits its representation of the domain. Third, the environment maintains the ground truth model of what the students are supposed to learn, and provides feedback to the agent on whether the student has taught the correct or incorrect information.

The **student** acts as a more knowledgeable peer, teaching the virtual agent through a chat interface. These student-teachers are responsible for choosing the relevant content to teach the agent, monitoring the agent's learning progress, and providing feedback to the agent (e.g., a high-five for a job well done) based on their understanding of its current state of knowledge and learning progress.

Unlike the tutoring agents in traditional pedagogical systems, the **conversational agent** in Curiosity Notebook takes the role of a less intelligent peer that can be taught through interactions in natural language. The teachable agent can exhibit different personality traits and use a variety of verbal interventions to improve different aspects of student learning by prompting them to teach in different ways. The design of our agent is inspired from the generic agent architecture for autonomous agents proposed by Arkin et al. [1]. It consists of the following five subsystems: perception, planning, knowledge, motor, and homeostatic control (Figure 2).

The components of this socio-technical system work together to automatically identify the conversational interventions that are helpful for long-term engagement and inducing the maximum Protégé effect from the learning experience.

### Design Objectives

In order for the Curiosity Notebook to work in real-world classroom environments, we have to take into account the following design objectives:

- **The Need to Provide Equal Access.** All members of a student group can spend equal time interacting with the agent, so that students don't dominate each other while teaching.
- **Flexibility for Different Agent Embodiment.** The conversational agent should be designed with flexibility in mind, and take different physical forms—e.g., as a chatbot, physical robot (e.g., humanoid), or virtual character—depending on the learning context.
- **The Need to Safeguard Student Confidence.** Since the agent can have different rates of learning, we have to ensure that students feel confident as a teacher, regardless of whether they are teaching a slow or fast learner.
- **Keeping Low Start-up Cost.** The Curiosity Notebook needs to provide an easy way for teachers to set up the agent and the teaching materials for any topic.
- **Effective Ways to Prompt for Teaching.** The teachable agent needs to be designed to prompt students to teach in a variety of ways, e.g., asking them to provide examples, repeat or rephrase previous lessons, provide rationale behind facts, etc.
- **A Believable Model of the Agent as a Learner.** To facilitate learning by teaching, the agent needs to behave like a real learner. It is therefore important to accurately and convincingly reveal the agent's current state of knowledge and learning progress while it is being taught.

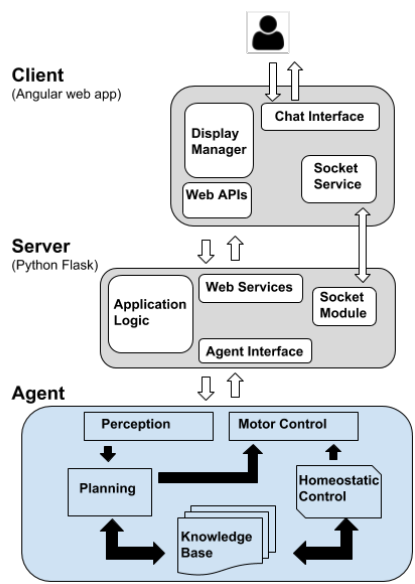

Figure 2: System Architecture of Curiosity Notebook

### User Study

We are working with students in an elementary school to study how we can use conversational agents to induce curiosity, and what characteristics of the agents are most helpful in realizing the Protégé effect.

## CONCLUSION

In this paper, we highlight the potential of conversational systems for educational purposes. We also list some design principles that can be incorporated in conversational agents to support learning by teaching. These principles are considered in the design of a learning environment called *Curiosity Notebook*, which supports conversational interaction between students and AI agents. User studies in this environment are necessary in order to elucidate the ways in which students interact with conversational agents in an educational context. Future research directions include further investigation into the possibilities of using conversational agents in education, and techniques for dynamically adapting the nature of the agent-student interaction to enhance learning and curiosity.